\documentclass[
    ,final            
  ]
  {aipproc}

\layoutstyle{6x9}

\begin{document}

\title{Gamma-Ray Bursts observed by INTEGRAL}

\author{S.Mereghetti}{
  address={INAF, IASF, via E.Bassini 15, Milano I-20133, Italy}
}



\begin{abstract}
During the first six months of operations, six Gamma Ray Bursts
(GRBs) have been detected
in the field of view of the INTEGRAL instruments and localized by
the  INTEGRAL Burst Alert System (IBAS): a software for
the automatic search of GRBs and the rapid distribution of their coordinates.
I describe the current  performances of IBAS and review the main
results obtained so far.
The coordinates of the latest burst localized by IBAS, GRB 031203, have been
distributed within 20 s from the burst onset and with an uncertainty
radius  of only 2.7 arcmin.
\end{abstract}

\maketitle


\section{Introduction}

The INTEGRAL satellite, devoted to high-resolution
imaging and spectroscopy in the hard X--ray / soft $\gamma$-ray energy range,
has been successfully launched on October 17, 2002.
The spacecraft carries two main instruments, SPI \cite{spi} and IBIS  \cite{ibis},
optimized respectively for spectroscopy and imaging   performances.
Both instruments provide images of the $\gamma$-ray sky in the $\sim$15 keV -- 10 MeV
energy range, using the coded aperture technique.
These two main  instruments  are complemented by an X-ray
monitor  (JEM-X \cite{jemx}), covering the 4-35 keV range,
and by an optical camera (OMC \cite{omc}) operating in the V band.
All the INTEGRAL instruments are co-aligned and provide a broad energy coverage
of the targets in the central part of the IBIS and SPI field of view.

The GRBs detected by IBIS can be localized with a precision of a
few  arc minutes   in near real time,
exploiting the continuous data link with the ground during INTEGRAL observations.
This task is performed by the INTEGRAL Burst Alert System
(IBAS, \cite{ibas}) described in Section 2.
The preliminary results on six GRB imaged by IBIS and SPI during the first year of
the mission are presented in Section 3.

\section{The INTEGRAL Burst Alert System}

Contrary to most other $\gamma$-ray astronomy satellites, no
on-board GRB triggering system is present on INTEGRAL.
Since the data are continuously transmitted to ground without important delays,
the search for GRB is done  at the  INTEGRAL Science Data
Centre (ISDC,  \cite{isdc}).
This offers the advantages of a larger computing power and more
flexibility for software and hardware upgrades, with respect to
systems operating on board satellites.

IBIS is the most appropriate instrument on board INTEGRAL for GRB localization,
thanks to its large field of view (29$^{\circ}\times$29$^{\circ}$) and
its capability to locate sources at the arcminute level.
Two different methods to look for GRBs,
using the data from the IBIS lower energy detector ISGRI
\cite{isgri}, are used in parallel in IBAS.

In the first method the ISGRI counting rate is monitored
to look for  significant excesses with respect to
a running average, in a way similar to traditional  triggering algorithms
used on-board previous satellites.
Several different integration times, ranging from 2 ms to 5.12 s, are sampled in parallel.
A rapid imaging analysis is performed only when
a significant counting rate excess is detected.
Imaging allows to eliminate many false triggers caused, e.g.
by instrumental effects or background
variations that do not produce a point source in the reconstructed sky images.
The second method is entirely based on imaging.
Images of the sky are continuously produced
(integration times from 10 to 40 s) and compared with the previous ones to search
for new sources.

An additional IBAS  program  is used to search for GRBs detected
by the Anti Coincidence System (ACS) of the SPI instrument  \cite{acs}.
No directional or energetic information is available. The resulting light curves
at 50 ms resolution are automatically posted on the ISDC WWW pages and are used for
Inter-Planetary Network GRB localizations \cite{ipn}.

The GRB positions derived by IBAS are
delivered via Internet to all the interested users.
For the GRBs detected with high significance,
this is done immediately by the software which sends \textit{Alert Packets}
using the UDP transport protocol.
In case of events with lower statistical significance, the alerts
are sent only to the members of the IBAS Localization Team, who perform
further analysis and, if the GRB is confirmed, can distribute its position
with an \textit{Off-line Alert Packet}.

\begin{figure}
  \includegraphics[height=.6\textheight,angle=90]{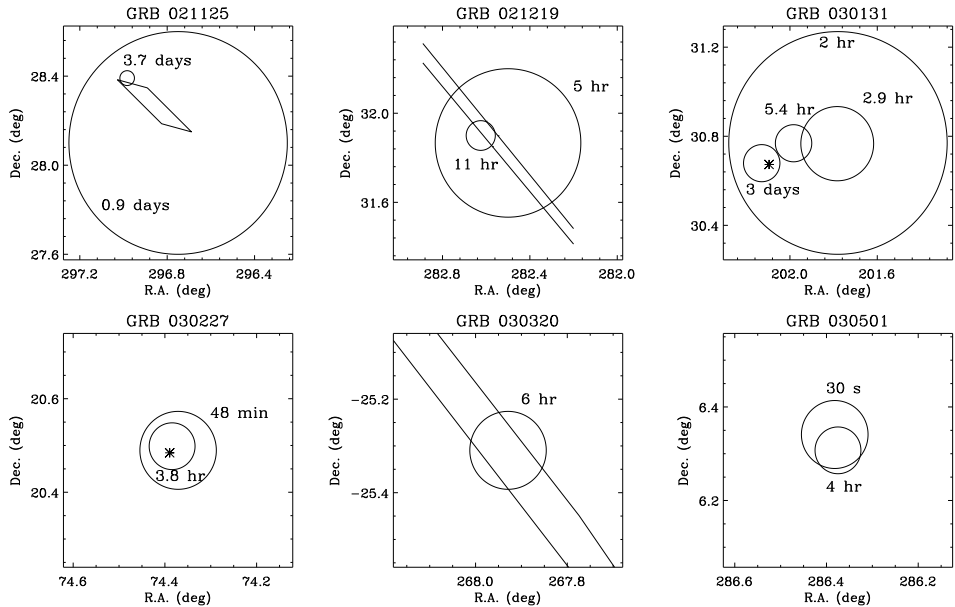}
  \caption{Error regions distributed for the six GRBs in the field of view of
      the INTEGRAL instruments, with the corresponding delays.
      The times refer to the public distribution of the GRB position.
      Note the different scale of the three upper (1$^{\circ}\times1^{\circ}$) and
      lower (0.5$^{\circ}\times0.5^{\circ}$)      panels.
      The parallelogram and the straight lines indicate error regions independently
      derived with the IPN (\cite{021125IPN,021219IPN,030320IPN}). The asterisks indicate the positions of the
      optical transients associated to GRB030131 (\cite{030131O})
      and GRB030227 (\cite{030227O})}
\end{figure}

\begin{table}
\begin{tabular}{lcccccc}
\hline
  & \tablehead{1}{r}{b}{Duration\\(s)}
  & \tablehead{1}{r}{b}{Distribution delay\\internal/public}
  & \tablehead{1}{r}{b}{Alert\\distribution}
  & \tablehead{1}{r}{b}{Peak Flux\\(ph cm$^{-2}$ s$^{-1}$)}
  & \tablehead{1}{r}{b}{Fluence\\(erg cm$^{-2}$)}
  & \tablehead{1}{r}{b}{Ref.}\\
\hline
021125  &  25  & -- / 0.9 days  & OFF & 22    &7.4$\times 10^{-6}$ & \cite{021125D,021125P}  \\
021219  &   6  & 10 s / 5 hr    & OFF & 3.7   &  9$\times 10^{-7}$ & \cite{021219D,021219P} \\
030131  &  150 & 21 s / 2 hr    & ON  & 1.9   &  7$\times 10^{-6}$ & \cite{030131D,030131P} \\
030227  &  20  & 35 s / 48 min  & OFF & 1.1   &7.5$\times 10^{-7}$ & \cite{030227D,030227P} \\
030320  &  50  & 12 s /  6 hr   & ON  & 5.7   &1.1$\times 10^{-5}$ & \cite{030320D,030320P} \\
030501  &  40  & 30 s / 30 s    & ON  & 2.7   &  3$\times 10^{-6}$ &  \cite{030501D,030501P} \\
\hline
\end{tabular}
\end{table}

The time delay in the automatic distribution of coordinates results from the sum of
several factors. There is a first delay of the data on board the satellite,
which is variable and  depends on the instrument.
In the case of IBIS/ISGRI data the average delay is about 5 s,
but it can be much longer for other instruments (e.g. approximately 20 s
on average for the SPI ACS data).
Signal propagation to the ground station is negligible (maximum $\sim$0.6 s),
but some time is required before the data are received at the ISDC.
This is on average 3 s when the ESA   ground station in Redu (Belgium)
is used, or 6 s when the NASA Goldstone ground station is used.
The time to detect the GRB depends on the algorithm which triggers.
The delay between the trigger time and the GRB onset
is of course dependent on the intensity and  time profile of the event.
The IBAS simultaneous sampling in different timescales should ensure
a minimum delay in most cases.
Thus, for GRBs lasting a few tens of second, IBAS can in principle distribute the alerts
with the position while the  GRB is still ongoing.

The OMC  covers only the central
5$^{\circ} \times$5$^{\circ}$  of the IBIS and SPI field of view.
During normal operations
only the data from a number of small pre-selected windows around sources of interest are
recorded and transmitted to the ground.
The  IBAS programs  check whether the  GRB position falls within the OMC field of view.
In such a case, the appropriate telecommand with the definition
of a new window centered on the interesting region is sent to the satellite.
The OMC observation will consist of several frames with integration times of 60 s
to permit variability studies and to increase the sensitivity for very intense but
short outbursts. The expected limiting magnitude
is of the order of V$\sim$14   (60 s at high Galactic latitude).

\begin{figure}
  \includegraphics[height=.6\textheight,angle=90]{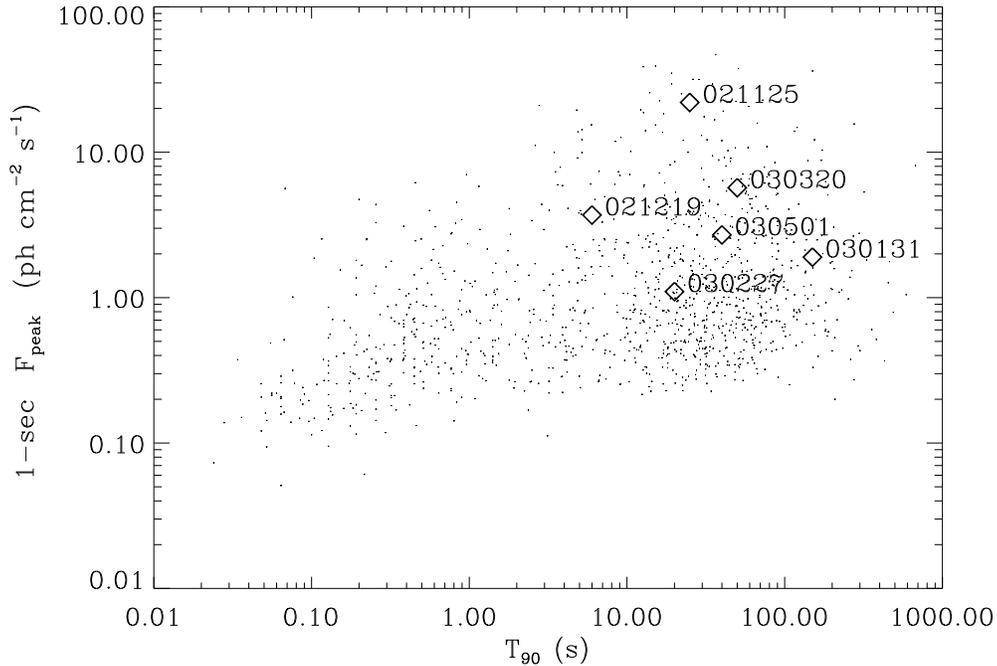}
  \caption{Peak flux versus duration for the six GRBs detected by INTEGRAL
  compared to the corresponding values of the BATSE catalog.}
\end{figure}

\section{Results}

IBAS has been running almost continuously since the launch of  INTEGRAL.
The first two months of operations were devoted to the tuning of the
IBAS parameters. Some changes in the
algorithms were also required to adapt them to the in-flight data characteristics.
Delivery of the \textit{Alert Packets} to the external clients started on January 17, 2003.
Since then it has always been enabled, except for the period from
February 12 to 28 (during calibration observations of the Crab Nebula),
and for a short interruption (4 hours) on April 23 (for hardware maintenance
reasons).

Six GRBs have been discovered in the field of view of IBIS during the first
six months of operations.
All of them were at relatively large off-axis angles,
outside the fields of view of the OMC and JEM-X instruments.
Two GRBs occurred during the initial  performance and verification phase
(GRB 021125 and GRB 021219), and  at the time of GRB 030227,
as mentioned above, the automatic distribution
of alerts to the external users was disabled.
Of the remaining three bursts, only GRB 030501 was detected at a significance
level high enough for automatic delivery of the coordinates.

The error regions derived for these six GRBs are shown in
Figure 1, with the corresponding time delays in the public distribution of the coordinates.
Note that at the beginning of the mission the in-flight
instrument misalignment was not calibrated yet.
Therefore,  conservative error radii as large as 20$'$ or 30$'$ were given.
The error regions obtained with the IPN,  and the coordinates
of the optical transients discovered for the two GRBs for which prompt observations
could be done, are also shown in the figure. Their agreement with the
INTEGRAL positions confirms that the IBAS localizations are correct.

The main properties of these six  GRBs  are summarized in Table 1.
All of them where of the long duration class. Figure 2 shows their peak flux
and T$_{90}$ duration compared to the corresponding values of the bursts in the
BATSE Catalog \cite{batse}.

The first burst to be imaged by INTEGRAL, GRB 021125 is also the only one
for which data from the high-energy IBIS/PICsIT \cite{picsit} detector were obtained.
In fact during this observation, PICsIT was operated in photon-by-photon mode
(in the  standard mode of operations PICsIT collects images integrated over
a few thousand seconds and does not have enough  time resolution for GRBs studies).
Thus the spectrum of GRB 021125 could be studied up to $\sim$500 keV \cite{021125P}.

The second burst in the IBIS field of view, GRB 021219,
was found and correctly localized by IBAS in real time \cite{021219D}.
The position  derived by the IBAS software within $\sim$10 s
had an accuracy of $\sim20'$. As mentioned above, this error was largely dominated
by systematic uncertainties present in the early phase of the mission.
In less than four hours the error region could be significantly reduced,
exploiting the presence of Cyg X-1 in the field of view of the same observation.
Time resolved spectroscopy of GRB 021219 indicated a clear hard to soft evolution:
the 15-200 keV spectrum  was fit by a single power law
with photon index evolving from  1.3 to 2.5 \cite{021219P}.

GRB 030131 lasted $\sim$150 s, but only the first 20 s were observed during a stable
pointing, after which the satellite started a slew to the next pointing direction.
Therefore, the signal to noise ratio of the trigger was not high enough for the
automatic alert distribution. The moving satellite aspect complicated the
analysis, resulting initially in  a wrong  localization (see Fig. 1).
Nevertheless, a faint optical transient could be identified \cite{030131O}.
This burst had a fluence  of 7$\times$10$^{-6}$ erg cm$^{-2}$  (20-200 keV) and an
average spectrum well described by a Band function with break energy
$E_{0}$=70$\pm$20 keV,
$\alpha$ = 1.4$\pm$0.2 and  $\beta$=3.0$\pm$1.0 \cite{030131P}.
Time resolved spectroscopy indicated also for this burst
a hard-to-soft evolution.

The quick localization \cite{030227D} obtained for GRB 030227 led to the discovery
of both its X--ray \cite{030227P} and optical afterglows \cite{030227O}.
This burst had a duration of about 20 s and a peak flux of
$\sim$1.1 photons cm$^{-2}$ s$^{-1}$ in the 20-200 keV energy range.
The spectrum was a  power law with
average photon index  $\sim$2 and some  evidence for a  hard to soft
evolution \cite{030227P}.
\textit{XMM-Newton}  started a Target of Opportunity Observation
only  8 hours after the GRB. The X--ray afterglow was discovered
with a 0.2-10 keV  flux decreasing as t$^{-1}$ from 1.3$\times$10$^{-12}$
to 5$\times$10$^{-13}$ erg cm$^{-2}$ s$^{-1}$.
The afterglow spectrum was well described by a power law
with photon index 1.94$\pm$0.05.
Interestingly,  a significant absorption in addition to the
Galactic value was required to fit the X--ray data \cite{030227P}.
The exact value of this intrinsic absorption depends on the (unknown) redshift,
but is in any case of the order of a few times   10$^{22}$ cm$^{-2}$.
This supports the scenarios involving the occurrence
of GRBs in regions of star formation.
Some evidence for an emission line at  1.67 keV,
which if attributed to Fe would imply a redshift $z\sim$3,
was also found in the   \textit{XMM-Newton} spectrum  \cite{030227P}.
Contrary to recent claims \cite{watson}, we find that  dividing the observation
in short time intervals, all the spectra are well fit by the  non-thermal power law
spectrum, without the need of emission lines from light elements.

The next burst, GRB 030320 demonstrated the IBAS capability to discover and
correctly locate GRBs even at very large off-axis angles \cite{030320P}.
The photons of this GRB, located more than 15$^{\circ}$
from the pointing direction, illuminated only a very small
fraction of the IBIS/ISGRI detector and only three  of the 19 SPI pixels.
This GRB was coming from a direction very close to that of the Galactic Center and
no  optical observations of its error region were reported.

The Labour Day burst, GRB 030501, is the one with the best combination of
speed and accuracy in localization.
Its coordinates with an uncertainty of only 4.4$'$
reached all the IBAS users 30 s after the beginning of the event.
Observations with robot telescopes   started while the gamma-ray emission
was still visible \cite{tarot}, but unfortunately
GRB 030501 was at low Galactic latitude, in a region of very
high interstellar absorption which hampered sensitive searches for counterparts.

Finally, thanks to the extension in the deadline for submission of these proceedings,
I can add the results on the latest burst located by IBAS, GRB 031203
\cite{D031203,D031203b}.
It triggered with high significance on several timescales,  setting a new record
in localization speed and accuracy. Its position with an error of 2.7 arcmin
was distributed by IBAS less than 20 s from the burst start time.

\section{Conclusions}

Thanks to the good imaging capabilities of the IBIS instrument and
the continuous contact with the ground
stations during the INTEGRAL observations,
IBAS represents a step forward compared to previous  GRB localization facilities.
As demonstrated by GRB 030501 and GRB 031203,
IBAS is currently able to provide small error regions
($\sim3-4'$ radius) within a few tens of seconds from the GRB onset.




\bibliographystyle{aipprocl} 

\bibliography{sample}


\IfFileExists{\jobname.bbl}{}
 {\typeout{}
  \typeout{******************************************}
  \typeout{** Please run "bibtex \jobname" to optain}
  \typeout{** the bibliography and then re-run LaTeX}
  \typeout{** twice to fix the references!}
  \typeout{******************************************}
  \typeout{}
 }

\end{document}